\documentclass[%
reprint,
superscriptaddress,
nofootinbib,
amsmath,amssymb,
aps,
prl,
]{revtex4-2}

\usepackage{times}
\usepackage{newtxmath}

\usepackage{graphicx}
\usepackage{mathrsfs}
\usepackage{bm}
\usepackage{xcolor}
\usepackage{hyperref}
\usepackage{wasysym}
\usepackage{physics}
\usepackage{orcidlink}
\definecolor{darkblue}{rgb}{0.0, 0.1, 0.7}
\hypersetup{
	colorlinks=true,
	linkcolor=darkblue,
	filecolor=darkblue,      
	urlcolor=darkblue,
	citecolor = darkblue,
	pdftitle={Separability of the motion of spinning test particles in curved space-time},
}

\usepackage{fancyhdr}
\begin{document}
	% Custom commands:
	\renewcommand{\d}{\mathrm{d}}
	\newcommand{\Tt}{\Tilde{t}}
	\renewcommand{\Tr}{\Tilde{r}}
	\newcommand{\Tz}{\Tilde{z}}
	\newcommand{\Tphi}{\Tilde{\phi}}
	\newcommand{\Tl}{\Tilde{\lambda}}
	
	\newcommand{\mytitle}[1]{\noindent
		\emph{#1.}--}
	
	\newcommand{\vs}[1]{\textcolor{blue}{VS: #1}}
	\newcommand{\vw}[1]{\textcolor{red}{VW: #1}}
	
	%\renewcommand{\textbf}[1]{{#1}}

%%%%%%%%%%%%%%%%%%%%%%%%%%%%%%%%%%%%%%%%%%%%%%%%%%%%%%%%%%%%%
\title{Separability of the motion of spinning test particles in curved space-time} 
%%%%%%%%%%%%%%%%%%%%%%%%%%%%%%%%%%%%%%%%%%%%%%%%%%%%%%%%%%%%%

\author{Vojt{\v e}ch Witzany \orcidlink{0000-0002-9209-5355}}
\email{vojtech.witzany@matfyz.cuni.cz}

\author{Viktor Skoup\'y \orcidlink{0000-0001-7475-5324}}
\email{viktor.skoupy@matfyz.cuni.cz}

\affiliation{Institute of Theoretical Physics, Faculty of Mathematics and Physics, Charles University, CZ-180 00 Prague, Czech Republic}

%%%%%%%%%%%%%%%%%%%%%%%%%%%%%%%%%%%%%%%%%%%%%%%%%%%%%%%%%%%%%
%%%%%%%%%%%%%%%%%%%%%%%%%%%%%%%%%%%%%%%%%%%%%%%%%%%%%%%%%%%%%
\begin{abstract}
    Solving for the motion of spinning test particles in curved spacetimes is important for modeling gravitational-wave inspirals of spinning compact binaries. We build a Hamiltonian formalism in worldline-adapted tetrads for the spinning test particle and formulate a corresponding Hamilton-Jacobi equation valid to linear order in spin. We prove that when the geodesic motion in a spacetime and the parallel transport along said geodesics are both separable, then so is the corresponding Hamilton-Jacobi equation. We illustrate this in black hole, plane wave, and cosmological spacetimes.
\end{abstract}
%%%%%%%%%%%%%%%%%%%%%%%%%%%%%%%%%%%%%%%%%%%%%%%%%%%%%%%%%%%%%
%%%%%%%%%%%%%%%%%%%%%%%%%%%%%%%%%%%%%%%%%%%%%%%%%%%%%%%%%%%%%
\maketitle
	
%%%%%%%%%%%%%%%%%%%%%%%%%%%%%%%%%%%%%%%%%%%%%%%%%%%%%%%%%%%%%
\mytitle{Introduction}
The motion of a spinning test body in a curved background is a classic problem of relativistic mechanics that finds use in modeling extreme-mass-ratio inspirals of rotating compact objects for future space-based detectors such as LISA and beyond \citep{LISA,LISAConsortiumWaveformWorkingGroup:2023arg}. Geodesic motion in black-hole space-times is separable and integrable thanks to the hidden symmetries encoded in Killing-Yano tensors \citep{Carter:1968ks,Frolov:2017kze}. Endowing the particle with spin generically spoils this separability and, moreover, forces one to confront the relativistic ambiguity in defining the body's center of mass -- the choice of a spin supplementary condition. This choice is the classical counterpart of the Newton-Wigner localization of relativistic particles \citep{pryce1948mass,Newton:1949cq,witzany2023spinning}, and it is what underlies the canonical, Newton-Wigner-type phase-space variables used also in the modeling of spinning compact binaries at comparable mass ratios \citep{Barausse:2009aa,Barausse:2009xi}.

In this Letter, we recast the linear-in-spin dynamics as a constrained Hamiltonian system in a worldline-adapted, phase-space-dependent frame, in which the supplementary condition becomes a simple Dirac constraint that generalizes the Newton-Wigner variables. We then show that a suitable \emph{non-minimal} choice of frame renders the Hamilton-Jacobi (HJ) equation of the spinning particle separable at linear order in spin precisely when the parallel transport of vectors along the background geodesics is itself separable. After establishing this general result, we illustrate it on three families of space-times---Kerr-NUT-(A)dS black holes, gravitational plane waves, and FLRW cosmologies. A companion paper provides further details on the formalism and develops it for higher-dimensional black hole spacetimes.

%%%%%%%%%%%%%%%%%%%%%%%%%%%%%%%%%%%%%%%%%%%%%%%%%%%%%%%%%%%%%
\mytitle{Notation}
We use geometrized units with $G=c=1$ and the mostly plus metric signature $(-,+,...,+)$. Space-time indices are denoted by lowercase Greek letters and Latin-letter indices $A,B,C,...=0,...,N-1$ denote tetrad components, where $N$ is the general space-time dimension. 

%%%%%%%%%%%%%%%%%%%%%%%%%%%%%%%%%%%%%%%%%%%%%%%%%%%%%%%%%%%%%
\mytitle{Canonical formalism for spinning particles} Consider the canonical action for a spinning body in curved space-time \citep{bailey1975lagrangian,hanson1974relativistic}
\begin{align}
    & S = \int \tilde{p}_\mu \dot{\tilde{x}}^\mu + \frac{1}{2} \tilde{S}_{AB} \tilde{\Omega}^{AB} - H(\tilde{p}_\mu,\tilde{x}^\mu,\tilde{S}_{AB}) \d \tau \,, \\
    & \tilde{\Omega}^{AB} \equiv \tilde{\Lambda}^A{}_{\hat{A}}\frac{\d \tilde{\Lambda}^{B \hat{A}}}{\d \tau} \,,
    \;\; \tilde{S}_{AB} \equiv 2\tilde{\Pi}_{A}{}^{\hat{A}} \tilde{\Lambda}_{B\hat{A}}
\end{align}
where $\tilde{x}^\mu$ is worldline coordinate and $\tilde{p}_\mu$ its canonically conjugate momentum, and $\tilde{\Pi}_A{}^{\hat{A}},\tilde{\Lambda}^A{}_{\hat{A}}$ are canonically conjugate variables parametrizing the frame orientation of the body with respect to the ambient spacetime frame. The hatted index is with respect to the body-fixed frame, whereas the non-hatted index with respect to some fixed background tetrad $\tilde{e}^\mu_A$. We have $\tilde{\Lambda}^A{}_{\hat{A}} \tilde{\Lambda}^B{}_{\hat{B}} \eta^{\hat{A}\hat{B}} = \eta^{AB}$, $\tilde{\Lambda}^A{}_{\hat{A}} \tilde{\Lambda}^B{}_{\hat{B}} \eta_{AB} = \eta_{\hat{A}\hat{B}}$. The spin tensor then has the Poisson bracket of the generators of the Lorentz group $\{\tilde{S}_{AB},\tilde{S}_{CD}\} =\eta_{AC}\tilde{S}_{BD} + \eta_{BD}\tilde{S}_{AC} - \eta_{AD}\tilde{S}_{BC} -  \eta_{BC}\tilde{S}_{AD}$. We assume bodies that do not have any directional features apart from their rotation, which is reflected by the fact that the Hamiltonian only depends on $\tilde{S}_{AB}$. Since in the vast majority of applications of interest, the coupling of the spin degree of freedom is small, we will treat it strictly perturbatively to linear order. We will use the minimal Hamiltonian
\begin{align}
    & H = \frac{1}{2 \mathcal{M}} g^{\mu\nu}(\tilde{x}) P_\mu P_\nu \,,  \label{eq:HminNT}
    & P_\mu \equiv \tilde{p}_\mu - \frac{1}{2} \tilde{e}_{\nu A;\mu} \tilde{e}^\nu_{B} \tilde{S}^{AB}\,,
\end{align}
where $P_\mu$ is known as the covariant momentum and $\mathcal{M}$ the particle mass such that $g^{\mu\nu} P_\mu P_\nu = -\mathcal{M}^2$ on shell. 
Then the variation of the actions is equivalent to the Mathisson-Papapetrou-Dixon equations \citep{Witzany:2018ahb}
\begin{align}\label{eq:MPD_full}
    \frac{\mathrm{D} P^\mu}{\dd \tau} &= -\frac{1}{2} R^{\mu}{}_{\nu\rho\sigma} \dot{\tilde{x}}^\nu \tilde{S}^{\rho\sigma} \,, 
    & \frac{\mathrm{D} \tilde{S}^{\mu\nu}}{\dd \tau} &= 0 + \mathcal{O}(\tilde{S}^2) \,.
\end{align}
In this form, there is unconstrained freedom in the choice of center of mass, which manifests in an unphysical degree of freedom in the spin tensor $\tilde{S}^{\mu\nu}$. This can be reduced using a Dirac constraint procedure corresponding, e.g., to the Tulczyjew-Dixon, or ``covariant'' supplementary condition $S^{\mu\nu}P_\nu =0$ \citep{tulczyjew1959motion,Dixon:1970I,Witzany:2018ahb}. Before doing so, however, we will choose a better basis of coordinates on the phase space.

%%%%%%%%%%%%%%%%%%%%%%%%%%%%%%%%%%%%%%%%%%%%%%%%%%%%%%%%%%%%%
\mytitle{Worldline adapted frame} The advantage of the coordinates $\tilde{p}_\mu, \tilde{x}^\mu$ and $\tilde{\Pi}_A{}^{\hat{A}},\tilde{\Lambda}^A{}_{\hat{A}}$ is that they are simple canonical conjugates. We can thus generate new canonical coordinates by a standard generating function \citep{goldstein1950classical}. Consider a generating function of the second kind $G(\tilde{x}^\mu,p_\nu,\tilde{\Lambda}^B{}_{\hat{A}},\Pi_{B}{}^{\hat{A}})$, which generates new coordinates $p_\mu, x^\mu$ and $\Pi_{A}{}^{\hat{A}},\Lambda^{A}{}_{\hat{A}}$ through the standard relations
\begin{align}
    \tilde{p}_\mu = \frac{\partial G}{\partial \tilde{x}^\mu}\,,\; 
    \;&\;
    x^\mu =\frac{\partial G}{\partial p_\mu}\,,\;
    \\ 
    \tilde{\Pi}_{B}{}^{\hat{A}} = \frac{\partial G}{\partial \tilde{\Lambda}^B{}_{\hat{A}}} \,, 
    \;&\;
    \Lambda^{B}{}_{\hat{A}} = \frac{\partial G}{\partial \Pi_{B}{}^{\hat A}}\,.
\end{align}
We describe a change from the background frame components to a generic, phase-space dependent frame by the generating function
\begin{align}
    G = \tilde{x}^\mu p_\mu + \tilde{\Lambda}^{C}{}_{\hat{A}} L^{B}{}_C(\tilde{x}^\mu,p_\nu) \Pi_{B}{}^{\hat A}\,,
\end{align}
where we assume that $L^{B}{}_C$ is a Lorentz transform at any point in the mixed phase-space $\tilde{x}^\mu,p_\nu$.  The spin then has the transform corresponding to a usual tetrad change
\begin{align}
    S_{AB} \equiv 2\Pi_{A}{}^{\hat{A}} \Lambda_{B\hat{A}} = \tilde{S}_{CD} L_{A}{}^C L_{B}{}^D \,, \label{eq:newS}
\end{align}
and, importantly, keeps the same Poisson bracket structure. 
Using this we can express the transformation of momentum as
\begin{align}
\begin{split}
    \tilde{p}_\mu & = p_\mu + \tilde{\Lambda}^{B}{}_{\hat{A}} \Pi_{C}{}^{\hat A} \frac{\partial L^{C}{}_B}{\partial \tilde{x}^\mu}
    \\
    & = p_\mu + \frac{1}{2} S_{BA} L^{A C} \frac{\partial L^{B}{}_C}{\partial \tilde{x}^\mu} \,.
\end{split}
\end{align}
When $L_{A}{}^B$ depends only on coordinates $\tilde{x}^\mu$ this reduces to well known formulas for background tetrad changes. However, in the general case we also have a coordinate shift of the form
\begin{align}
\begin{split}
    {x}^\mu & = \tilde{x}^\mu + \tilde{\Lambda}^{B}{}_{\hat{A}} \Pi_{C}{}^{\hat A} \frac{\partial L^{C}{}_B}{\partial p_\mu}
    \\
    & =  \tilde{x}^\mu + \frac{1}{2} S_{BA} L^{A C} \frac{\partial L^{B}{}_C}{\partial p_\mu}\,. \label{eq:newx}
\end{split}
\end{align}
These relations can be easily inverted at $\mathcal{O}(S)$ by iterating from identity. To generate a relevant $L^A{}_B$, it is easiest to pick a phase-space dependent, or \textit{worldline-adapted} tetrad $e^\mu_{A}(x^\mu,p_\nu)$ and $L_{A}{}^B = e^\mu_{A}(\tilde{x}^\mu,p_\nu) \tilde{e}_\mu^B(\tilde{x}^\mu)$. More details can be found in the End Matter. Up to higher-order terms, the minimal Hamiltonian from Eq. \eqref{eq:HminNT} transforms into
\begin{align}
    H = \frac{1}{2 \mathcal M}  g^{\mu\nu}(x^\alpha) p_\mu \left(p_\nu - \mathcal{D}_\nu e^\kappa_A e_{B\kappa} S^{AB} \right) \label{eq:Hmin}
\end{align}
where we defined the covariant derivative on the phase space
\begin{equation}
    \mathcal{D}_\mu e^\kappa_A = e^\kappa_{A,\mu} + \Gamma^\kappa{}_{\mu\lambda} e^\lambda_A + \Gamma^\nu{}_{\mu\rho} p_\nu \pdv{e^\kappa_A}{p_\rho}\,.
\end{equation}
This formalism opens the possibility to search for advantageous choices of adapted frames $e^\mu_{A}(x^\mu,p_\nu)$. 

%%%%%%%%%%%%%%%%%%%%%%%%%%%%%%%%%%%%%%%%%%%%%%%%%%%%%%%%%%%%%
\mytitle{Dirac constraint in momentum frame} Consider an adapted tetrad $e^\mu_{A}(x^\mu,p_\nu)$ that takes $e^\mu_{0} = p^\mu/\sqrt{-p_\mu p^\mu}$. In this frame, the Tulczyjew-Dixon condition has the simple form $S_{0B} =0 + \mathcal{O}(S^2)$. Since $S_{0B}$ only has a non-zero Poisson bracket with itself, this provides an easy way to introduce the Dirac bracket corresponding to this constraint \citep{Henneaux:1992ig}; simply ignore the brackets involving $S_{0B}$ and work only with the Hamiltonian and other phase space functions expressed strictly on the constraint hypersurface $S_{0B} =0$. The transformation of the minimal Hamiltonian \eqref{eq:Hmin} then yields to leading order
\begin{align}
    & H = \frac{1}{2 \mathcal M}  g^{\mu\nu}(x^\alpha) p_\mu \left(p_\nu - \mathcal{D}_\nu e^\kappa_{I} e_{J\kappa} S^{IJ} \right) \,, \label{eq:Hred}
\end{align}
where we have introduced the spatial indices $I,J = 1,...,N-1$.
The coordinates $x^\mu, p_\nu$ are already canonically conjugate and commute with $S_{IJ}$. The $S^{IJ}$ sector has the Poisson bracket is the same as the $\mathfrak{so}(N-1)$ Lie bracket and it can thus be covered by an appropriate set of Darboux variables. The algebra $\mathfrak{so}(N-1)$ has $[(N-1)/2]$ Casimirs and $(N-1)(N-2)/2$ elements, where $[]$ is the integer part \citep{bincer2013lie}. This yields $((N-1)(N-2)/2-[(N-1)/2])/2$ of pairs of canonical coordinates or Darboux variables. We will present the Darboux variables for generic $N$ in the companion paper. In $N=4$ we can use the usual canonical coordinates $\pi_\phi, \phi$ such that $S^{12} \!= \pi_{\phi}$, $\, S^{23}\! = \sqrt{S^2 - \pi_\phi^2} \cos \phi$, $S^{31}\! = \sqrt{S^2 - \pi_\phi^2} \sin \phi$,
and the rest can be recovered by the antisymmetry of $S^{IJ}$. This represents the full reduction of phase space of a spinning test particle in a curved space-time that now depends only on the choice of the $N-1$ comoving basis legs $e^\mu_{I} (x^\nu,p_\kappa)$. For example, by minimally boosting from the background tetrad frame $\tilde{e}^\mu_A$, we obtain a formalism equivalent to the Newton-Wigner variables of \citet{Barausse:2009aa} (for more details, see the End Matter). However, we will show that it is advantageous to work with \textit{non-minimal} choices of tetrads that separate parallel transport along geodesics.

%%%%%%%%%%%%%%%%%%%%%%%%%%%%%%%%%%%%%%%%%%%%%%%%%%%%%%%%%%%%%
\mytitle{Separability of parallel transport} Consider parallel transport along a geodesic $x^\mu(\tau)$ parametrised by proper time $\tau$ and a worldline-adapted tetrad that expressed as a function of position and covariant components of four-velocity, $e^\mu_A (x^\kappa,u_\mu)$, $u_\mu \equiv g_{\mu\nu}\mathrm{d}x^\nu/\mathrm{d}\tau$. We take $e^\mu_0 = g^{\mu\nu} u_\nu/\sqrt{-g^{\alpha \beta}u_\alpha u_\beta}$ and the rest of the legs as generic functions $e^\mu_I (x^\kappa,u_\nu)$ fulfilling orthogonality relations both on and off-shell (that is, even for $u_\mu$ that are not normalized). Then with respect to this tetrad, the zeroth component of any parallel-transported vector $V^\mu$ is transported trivially, $\d V^0/\d\tau =0$. The rest of the components can be transformed into the form
\begin{align}
    & \frac{\d V^I}{\d \tau} =-\left(\frac{\mathrm{d}e^\mu_J}{\mathrm{d} \tau} e_\mu^I + \Gamma^\mu{}_{\nu \kappa} e^\nu_{J} \dot{x}^\kappa e^I_\mu\right) V^J = \omega^I{}_J V^J\,,\;\; 
    \\ 
    & \omega^I{}_J\equiv -\left( \frac{\partial e^\mu_J}{\partial x^\kappa} e^I_\mu + \Gamma^\mu{}_{\nu \kappa} e^\nu_{J} e^I_\mu + \Gamma^\lambda{}_{\kappa \rho} \frac{\partial e^\mu_J}{\partial u_\rho} e^I_\mu u_\lambda \right) g^{\kappa \gamma} u_\gamma\,, \label{eq:omega}
\end{align}
where in the last equality we used the geodesic equation to reexpress $\mathrm{d} u_\rho/\mathrm{d} \tau$. The real $N-1 \times N-1$ skew-symmetric transport matrix $\omega^I{}_J$ is only dependent on position and four-velocity $x^\mu,u_\rho$, which induces dependence on proper time $\tau$ along the worldline. At every $\tau$, it has $[(N-1)/2]$ eigenplanes. These eigenplanes are not constant as a function of time, which makes the general transport problem quite complex. If the planes are independent of $\tau$ in a given tetrad, we will call parallel transport in this tetrad \textit{torus-reducible}. This can be also expressed as the condition that the precession matrices at different times commute, $\omega^I{}_J(\tau)\omega^J{}_K(\tau') = \omega^I{}_J(\tau')\omega^J{}_K(\tau)$ for any $\tau,\tau'$. Without loss of generality, we can then introduce a global ($\tau$-independent) transform of the tetrad frame that rotates the invariant planes into the adjacent pairs of tetrad directions $\{I,J\}= \{2k-1,2k\}$, $k =1,...,[(N-1)/2]$. Additionally, for even $N$, the direction corresponding to the $N-1$-th tetrad leg does not fall into any of the planes and is invariant/parallel transported on its own. The transport problem is then solved by the equations 
\begin{align}
    & V^{2k-1} = V^{2k-1}_{(0)} \cos \psi_{(k)} + V^{2k}_{(0)} \sin \psi_{(k)}\,,
    \\
    & V^{2k} =  - V^{2k-1}_{(0)} \sin \psi_{(k)} + V^{2k}_{(0)} \cos \psi_{(k)}\,, 
    \\
    & \psi_{(k)} = \int \omega^{2k-1}{}_{2k}(\tau) \d \tau \,, \label{eq:psiInt}
\end{align}
where $V^{2k-1}_{(0)},V^{2k}_{(0)}$ are integration constants, and $\psi_{(k)}$ is the precession phase within the invariant planes. In even $N$ we additionally have the trivially transported component $V^{N-1} = V_{(0)}^{N-1}$. In other words, torus-reducible parallel transport can be characterized as independent pairwise precessions that occur on the $k$-torus $\psi_{(k)} \in (0,2\pi], k=1,...,[(N-1)/2]$.

While torus reducibility is convenient, integrating the $\psi_{(k)}$ integrals in Eq. \eqref{eq:psiInt} may still be challenging. Non-trivial cases further require separability. We call geodesic motion separable if there exists a special set of separation space-time coordinates $\hat{x}^{\mu}$ such that the HJ equation for a geodesic,
\begin{align}
    g^{\mu\nu} W_{(\mathrm{g}),\mu} W_{(\mathrm{g}),\nu} = - \mathcal{M}^2\,, \label{eq:HJgeo}
\end{align}
has a separable solution in the form
\begin{align}
    W_{(\mathrm{g})} = \sum_\mu w_{(\mu)}(\hat{x}^\mu;C_a)\,,
\end{align}
where $C_a$ is some set of separation constants and $w_{(\mu)}$ are functions of a single coordinate $\hat{x}^\mu$. We can then extend the parallel transport along a single geodesic to parallel transport along the whole constant-$C_a$ congruence by noting that $\d \hat{x}^\mu/\d \tau = g^{\mu\nu}\partial_\nu w_{(\nu)}/\mathcal{M}$. The torus reducible parallel-transport equations then can be locally expressed as a set of partial differential equations for $k$ scalar fields
\begin{align}
    \frac{\partial \psi_{(k)}}{\partial \hat{x}^\kappa} \frac{\partial_\nu w_{(\nu)}}{\mathcal{M}} g^{\kappa\nu} = \omega^{2k-1}{}_{2k}\left(\hat{x}^\mu,\partial_\kappa w_{(\kappa)}/\mathcal{M}\right) \,. 
\end{align}
We say that parallel transport along geodesics is separable if this equation is additively separable. In other words, we say that parallel transport is separable when it is torus reducible and we can re-express the quadrature in equation \eqref{eq:psiInt} as a set of separate quadratures in $\hat{x}^\mu$ in the form
\begin{align}
    \psi_{(k)}(\hat{x}^\nu;C_a) = \sum_\mu \int \Psi_{(k|\mu)}(\hat{x}^\mu;C_a) \mathrm{d} \hat{x}^\mu \,. \label{eq:psiSepInt}
\end{align}
 In $N=4$ there is only a single precession angle and the route to separation is usually simple and involving a single separation factor for all equations. However, in $N>4$ we will show in the companion paper that the geodesic and precession equations are generally ``multiseparable'' with no single separation factor.

%%%%%%%%%%%%%%%%%%%%%%%%%%%%%%%%%%%%%%%%%%%%%%%%%%%%%%%%%%%%%
\mytitle{Separability of HJ equation} Consider the HJ equation for a spinning particle corresponding to the Hamiltonian \eqref{eq:Hred}
\begin{align}
    g^{\mu\nu}W_{,\mu}W_{,\nu} - S^{IJ} g^{\mu\nu} W_{,\mu} \left(\mathcal{D}_\nu e^\kappa_{I} e_{J\kappa}\right)_{p_\mu \to W_{,\mu}}  = -\mathcal{M}^2 \,,
\end{align}
where the characteristic function $W$ is now a function of space-time coordinates and canonical coordinates on the $S^{IJ}$ spin sector, which we denote collectively by $\phi$. Additionally, the connection term $(\mathcal{D}_\nu e^\kappa_{I} e_{J\kappa})$ is evaluated by first applying the definition of the covariant phase-space derivative, and only then substituting $p_\mu \to W_{,\mu}$. Furthermore, $S^{IJ}$ is expressed in canonical coordinates $\phi$ and the canonical momenta are replaced by derivatives of $W$ with respect to conjugate variables, $\pi \to W_{,\phi}$. 

We search for a perturbative solution of the form $W = W_{(0)}(x^\mu) + W_{(1)}(x^\mu,\phi)$, where $W_{(1)}$ is $\mathcal{O}(S)$ and we neglect $\mathcal{O}(S^2)$ terms. We obtain
\begin{align}
    & g^{\mu\nu} W_{(0),\mu} W_{(0),\nu} = - \mathcal{M}^2\,, \label{eq:HJ0}
    \\
    & g^{\mu\nu} W_{(0),\mu} W_{(1),\nu} = \frac{1}{2} S^{IJ} g^{\mu\nu} W_{(0),\mu}   \left(\mathcal{D}_\nu e^\kappa_{I} e_{J\kappa}\right)_{p_\lambda \to W_{(0),\lambda}}. \label{eq:HJ1}
\end{align}
It is useful to reexpress
\begin{align}
    & \frac{1}{\mathcal{M}}g^{\mu\nu} W_{(0),\mu}   \left(\mathcal{D}_\nu e^\kappa_{I} e_{J\kappa}\right)_{p_\mu \to W_{(0),\mu}} = \omega_{IJ}(x^\lambda,W_{(0),\kappa}/\mathcal{M})\,,
\end{align}
where we mean the strict functional expression for $\omega_{IJ}(x^\lambda,u_\kappa)$ in equation \eqref{eq:omega}. 

So far, this is a statement about the relationship of parallel transport and the HJ equation of spinning particles in general space-times. However, let us further assume that the corresponding space-time has torus-reducible parallel transport along geodesics. In that case, only the $I,J=2k-1,2k$ components of $S_{IJ}$ (and skew symmetric $S_{JI} = -S_{IJ}$) appear in the HJ equation. In other words, each unordered pair of indices does not share any element with any other unordered pair appearing in the equation. The bracket $\{S_{IJ},S_{KL}\} =\delta_{IK}S_{JL} +\delta_{JL}S_{IK} - \delta_{JK}S_{IL} - \delta_{IL}S_{JK}$ then vanishes for all such $[(N-1)/2]$ components. This implies that these $[(N-1)/2]$ components can always be used as a subset of canonical momenta on the $\mathfrak{so}(N-1)$ phase space, $\pi_{(k)} = S_{2k-1 \,2k}$ for $k=1,...,[(N-1)/2]$. For $N>5$, one has to complete the set by additional uncoupled momenta $\pi_{(k)}, k= [(N-1)/2]+1,...,((N-1)(N-2)/2-[(N-1)/2])/2$ to build the full set of canonical coordinate pairs. Now, denoting the canonically conjugate angles as $\phi_{(k)}$, we finally have $S_{2k-1 \,2k} \to \partial W_{(1)}/\partial \phi_{(k)}$ for $k=1,...,[(N-1)/2]$ in the perturbative HJ equation, or
\begin{align}
    \begin{split}
        & \frac{1}{2\mathcal{M}} S^{IJ} g^{\mu\nu} W_{(0),\mu}   \left(\mathcal{D}_\nu e^\kappa_{I} e_{J\kappa}\right)_{p_\lambda \to W_{(0),\lambda}} =
        \\
        &
         \!\!\sum_{k=1}^{[(N-1)/2]}\!\!\omega_{2k-1\,2k}(x^\lambda,W_{(0),\kappa}/\mathcal{M} ) \frac{\partial W_{(1)}}{\partial \phi_{(k)}} \label{eq:conTR} \,.
    \end{split}
\end{align}
Since all $\phi_{(k)}$ are obviously cyclical coordinates, we can use a separation Ansatz $\partial W_{(1)}/\partial \phi_{(k)} = \sigma_{(k)}$, where $\sigma_{(k)}$ are separation constants. For $k=1,...,[(N-1)/2]$ these new separation constants govern the deviation from the geodesic HJ equation; for higher $k$ the separation constants are redundant and correspond to static degrees of freedom of the spin tensor.

This is an elegant result on its own, when parallel transport is torus-reducible along geodesics, the orbital sector of the HJ equation decouples from the spin degrees of freedom and only depends on a set of integrals of motion corresponding to $S_{2k-1 \,2k}$ projections of the spin tensor in the adapted tetrad. However, let us further assume that the geodesic motion and parallel transport are also separable and transform from generic coordinates to separation coordinates $x^\mu \to \hat{x}^\mu$. Then we can compare equations \eqref{eq:HJgeo} through \eqref{eq:psiSepInt} with equations \eqref{eq:HJ0}, \eqref{eq:HJ1} and \eqref{eq:conTR} to realize that we can express the perturbative solution of the HJ equation as
\begin{align}
    & W_{({0})} = W_{(g)} = \sum_\mu w_{(\mu)}(\hat{x}^\mu;C_a)\,, \\
    & W_{(1)} = \!\!\!\sum_{k=1}^{[(N-1)/2]}\!\!\! \sigma_{(k)} \left(\phi_{(k)}+ \sum_\mu \int \Psi_{(k|\mu)}(\hat{x}^\mu;C_a) \mathrm{d} \hat{x}^\mu \right),
\end{align}
where we keep the notation for the separation functions $C_a$ the same as in the geodesic case even though their interpretation will not be the same under perturbation.
In conclusion, this means that \textit{the separable solution of parallel transport of ``test'' tensors along the worldline automatically provides all the information to separate the perturbation caused by a ``non-test'' spin tensor perturbing the worldline through spin-curvature coupling}. We now consider some $N=4$ examples. Throughout these examples we express results using four-velocity $u_\mu$; the canonical, full-mass expressions are easily restored by the substitution $u_\mu\to p_\mu/\mathcal{M}$, as detailed in the End Matter.

%%%%%%%%%%%%%%%%%%%%%%%%%%%%%%%%%%%%%%%%%%%%%%%%%%%%%%%%%%%%%
\mytitle{Black hole space-times} Consider Kerr-NUT-(A)dS space-times (also known as the Carter family of space-times) described by the metric \citep{Carter:1968ks,griffiths2009exact}
\begin{align}
    \begin{split}
        ds^2
        = 
        & -\frac{\Delta_r}{\rho^2}\!\left(dt - \bigl(a\sin^2\!\theta + 2n(1-\cos\theta)\bigr)\,d\varphi\right)^{\!2}
        \\
        &  + \frac{\Delta_\theta\sin^2\!\theta}{\rho^2}\!\left(a\,dt - \bigl(r^2+(a+n)^2\bigr)\,d\varphi\right)^{\!2}
        \\
        &
        + \frac{\rho^2}{\Delta_r}\,dr^2
        + \frac{\rho^2}{\Delta_\theta}\,d\theta^2,
    \end{split} \label{eq:CarterMetric}
\end{align}
where $\rho^2 = r^2 + (n + a\cos\theta)^2$ and $\Delta_r(r)$ and $\Delta_\theta(\theta)$ are metric functions given in the End Matter that depend on the black hole mass $M$, spin $a$, NUT charge $n$, cosmological constant $\Lambda$, and electric/magnetic charges $e,g$. Interestingly, the results presented here apply to the entire class of ``off-shell'' Carter metrics \citep{Carter:1968ks}, where the functions $\Delta_r(r)$ and $\Delta_\theta(\theta)$ are left arbitrary.
The Misner string is on the half-axis $\theta=\pi$ and the metric is regular at $\theta=0$. The geodesic HJ equation is well known to separate in this space-time as \citep{Carter:1968ks}
\begin{align}
    \begin{split}
        W_{(g)} = 
        & -E(t-t_0) + L(\varphi-\varphi_0) \pm \int \frac{\sqrt{R(r;E,L,K)}}{\Delta_r}\,\d r 
        \\ 
        & \pm \int \frac{\sqrt{\Theta(\theta;E,L,K)}}{\Delta_\theta}\,\d\theta \,,
    \end{split}
\end{align}
where $E,L,K$ are separation constants, and the full expression for the functions $R, \Theta$ is given in the End Matter.
This space-time possesses a Killing-Yano tensor $Y_{\mu\nu} = - Y_{\nu\mu}$, $Y_{\mu\nu;\kappa} = -Y_{\mu\kappa;\nu}$ of the form 
\begin{align}
    \begin{split}
        & Y_{\mu\nu}\,dx^\mu \wedge dx^\nu
        = 
        - r\sin\theta\,d\theta \wedge
        \bigl(a\,dt - [\,r^2+(a+n)^2\,]\,d\varphi\bigr)
        \\ & +
        (n + a\cos\theta)\,dr \wedge
        \bigl(dt - [\,a\sin^2\!\theta + 2n(1-\cos\theta)\,]\,d\varphi\bigr).
    \end{split}
\end{align}
As introduced in Ref.~\cite{Witzany:2019} one can iterate powers of this tensor on $u_\mu$ and orthogonalize to obtain a worldline-adapted tetrad generalizing the Marck tetrad \cite{Marck:1983}; the explicit legs are given in the End Matter. By construction, $e_{0\mu} = u_\mu$ and $e_{3\mu} \propto Y_{\mu}{}^\nu u_{\nu}$ are parallel transported along geodesics. We then get only a single non-trivial component of the transport matrix in this tetrad frame, which is \citep[cf.][]{Connel:2008}
\begin{align}
\begin{split}
    \omega_{12} = -\omega_{21} =
    & \frac{\sqrt{K}}{\rho^2}
    \Bigg(
    \frac{E\bigl(r^2 + (a+n)^2\bigr) - aL}{K + r^2}
    \\
    &
    +
    \frac{a\bigl(L - aE\sin^2\!\theta\bigr) - 2nEa\,(1-\cos\theta)}{K - (n+a\cos\theta)^2}
    \Bigg)\,.
\end{split}
\end{align}
Here $E\equiv-u_t$, $L\equiv u_\varphi$, and $K\equiv K^{\mu\nu}u_\mu u_\nu$ are the energy, azimuthal angular momentum, and Carter constant of the geodesic. Since $[(N-1)/2]=1$, there is a single precession angle $\psi =\psi_{(1)}$ with a solution
\begin{align}
\begin{split}
    \psi = & \int  \sqrt{K}\,\frac{E\bigl(r^2 + (a+n)^2\bigr) - aL}{K + r^2}\, \frac{d r}{\sqrt{R}}
    \\ & + \int \sqrt{K}\,\frac{a\bigl(L - aE\sin^2\!\theta\bigr) - 2nEa\,(1-\cos\theta)}{K - (n+a\cos\theta)^2}\, \frac{d \theta}{\sqrt{\Theta}}\,.
\end{split}
\end{align}
This allows us to express the perturbation to the Hamilton-Jacobi equation of a spinning test particle as $W_{(1)} = \sigma\,(\phi + \psi(r,\theta;E,L,K))\,,$
where $\sigma =\sigma_{(1)} = S_{12}$ is the single separation constant in the spin sector.

%%%%%%%%%%%%%%%%%%%%%%%%%%%%%%%%%%%%%%%%%%%%%%%%%%%%%%%%%%%%%
\mytitle{Plane waves}
%%%%%%%%%%%%%%%%%%%%%%%%%%%%%%%%%%%%%%%%%%%%%%%%%%%%%%%%%%%%%
Consider a plane-wave space-time in Rosen coordinates $(U,V,x,y)$ \citep{Rosen:1937,Bondi:1958aj},
\begin{align}
    \begin{split}
        \d s^2 = & 2\,\d U\,\d V + e^{2\beta}\Big[\,e^{2\gamma}\big(\cos\alpha\,\d x+\sin\alpha\,\d y\big)^2 
        \\
        & + e^{-2\gamma}\big({-}\sin\alpha\,\d x+\cos\alpha\,\d y\big)^2\Big]\,,
    \end{split}
\end{align}
with conformal factor $\beta(U)$, shear $\gamma(U)$, and polarization angle $\alpha(U)$.
The spacetime carries a covariantly constant null vector $\ell^\mu = \delta^\mu_V$, $\nabla_\mu \ell^\nu = 0$. Since $V$, $x$, and $y$ are cyclic, the geodesic HJ equation \eqref{eq:HJgeo} separates immediately,
\begin{align}
    \begin{split}
        W_{(0)} = &p_V V + p_x\, x + p_y\, y 
        \\
        & - \frac{1}{2 p_V}\!\int\!\Big[1 + e^{-2\beta}\big(e^{-2\gamma}p_\shortparallel^2 + e^{2\gamma}p_\perp^2\big)\Big]\,\d U\,,
    \end{split}
\end{align}
where $p_\shortparallel \equiv \cos\alpha\,p_x+\sin\alpha\,p_y$ and $p_\perp \equiv -\sin\alpha\,p_x+\cos\alpha\,p_y$, with separation constants $p_V, p_x, p_y$. Since $\d U/\d\tau = u^U = p_V$ is constant, the coordinate $U$ is itself an affine parameter along any geodesic.
To adapt a tetrad we take, in addition to $e^\mu_0 = g^{\mu\nu}u_\nu/\sqrt{-u_\kappa u^\kappa}$, the longitudinal leg built directly from the covariantly constant null vector,
\begin{align}
    & e^\mu_3 = \frac{\ell^\mu + (\ell_\nu e_0^\nu)\,e_0^\mu}{\lvert \ell_\nu e_0^\nu\rvert}\,,
\end{align}
and two transverse legs $e_1,e_2$ by orthogonalizing along the principal directions of the wave, given explicitly in the End Matter. By construction $e_{0\mu}$ and $e_{3\mu}$ are parallel transported along geodesics -- the former by the geodesic equation, the latter because $\ell^\mu$ is covariantly constant and $\ell_\nu e_0^\nu$ is conserved. The principal axes rotate at the rate $\d\alpha/\d U$, and the remaining precession-matrix component reads
\begin{align}
    \omega_{12} = -\omega_{21} = p_V\,\cosh 2\gamma \,\frac{\d \alpha}{\d U}\,,
\end{align}
which depends on $U$ alone. The single precession angle then reduces to a quadrature over $U$,
\begin{align}
    \psi(U) = \int \cosh 2\gamma\,\frac{\d \alpha}{\d U}\,\d U\,.
\end{align}
The spin correction to the characteristic function is then $W_{(1)} = \sigma(\phi +\psi(U))$	where $\sigma = S_{12}$ is again the single spin separation constant.
	
%%%%%%%%%%%%%%%%%%%%%%%%%%%%%%%%%%%%%%%%%%%%%%%%%%%%%%%%%%%%%
\mytitle{Cosmological space-times} %%%%%%%%%%%%%%%%%%%%%%%%%%%%%%%%%%%%%%%%%%%%%%%%%%%%%%%%%%%%%
%%%%%%%%%%%%%%%%%%%%%%%%%%%%%%%%%%%%%%%%%%%%%%%%%%%%%%%%%%%%%
As an example where the result turns very simple, consider a generic Friedmann-Lemaitre-Robertson-Walker (FLRW) space-time in $N=4$,
\begin{align}
    & \d s^2 = -\d t^2 + a(t)^2\,\gamma_{ij}\,\d x^i \d x^j \,,
    \\
    & \gamma_{ij}\,\d x^i \d x^j = \d\chi^2 + S_k(\chi)^2\big(\d\theta^2 + \sin^2\!\theta\,\d\varphi^2\big)\,,
\end{align}
where $\gamma_{ij}$ is the maximally symmetric metric of the spatial slices and $S_k(\chi) = \sin\chi,\,\chi,\,\sinh\chi$ for spatial curvature $k = +1,\,0,\,-1$. The six Killing vectors of $\gamma_{ij}$ are also Killing vectors of the full metric; along any geodesic the comoving momentum magnitude $\gamma^{ij}u_i u_j$ is conserved and the spatial projection of the worldline is a geodesic of $\gamma_{ij}$. The geodesic HJ equation \eqref{eq:HJgeo} therefore separates as
\begin{align}
    \begin{split}
        W_{(0)} = & \pm\!\int\!\sqrt{1 + P^2/a(t)^2}\,\d t  \pm\!\int\!\sqrt{Q - \tfrac{L^2}{\sin^2\!\theta}}\,\d\theta
        \\
        & + L\varphi \pm\!\int\!\sqrt{P^2 - \tfrac{Q}{S_k(\chi)^2}}\,\d\chi\,,
    \end{split}
\end{align}
with separation constants $P^2,Q,L$, where $P^2$ is the conserved value of $\gamma^{ij}W_{(0),i}W_{(0),j}$. An adapted tetrad can be constructed by orthogonalizing the momentum, the spatial direction of the momentum, and the radial direction $\rho_\mu = \partial_\mu \chi$ (see the End Matter). Interestingly, this entire tetrad is parallel transported and the transport matrix vanishes identically, $\omega_{IJ}=0$. The spin correction then collapses to the bare cyclic term $W_{(1)}=\sigma\,\phi$: the spin sector decouples completely and the worldline receives no $\mathcal{O}(S)$ correction in this set of phase-space variables.

%%%%%%%%%%%%%%%%%%%%%%%%%%%%%%%%%%%%%%%%%%%%%%%%%%%%%%%%%%%%%
\mytitle{Discussion} 
%%%%%%%%%%%%%%%%%%%%%%%%%%%%%%%%%%%%%%%%%%%%%%%%%%%%%%%%%%%%%
One of us \citep{Witzany:2019} has previously derived similar results in Kerr space-time as discussed here, but they did not rely on the adapted Marck tetrad but on one that agreed with it ``by coincidence'' on the space-time background. This led to divergences in the perturbative solution of the characteristic function that complicated practical implementations \citep{Piovano:2024}, and had to be side-stepped by subtle analytical continuation arguments to derive actions \citep{Witzany:2024ttz}. It was assumed that the analytical continuation referred to some unknown phase space coordinates in which characteristic function is regular and manifestly separable. These formerly implicit variables are the ones we now explicitly construct through relations \eqref{eq:newS}-\eqref{eq:newx}. Similarly, the related action-angle variables are those that appear in the construction of the gravitational-wave flux of the R{\"udiger}-Carter constant in Refs \citep{Grant:2024ivt,Mathews:2025nyb}. We also verified that in the Kerr limit our formalism provides equivalent predictions as previous works \citep{Drummond:2022a,Drummond:2022b,Piovano:2024,Skoupy:2025}. We invite the reader to read our companion paper where we provide further details and apply this formalism to black hole space-times in higher dimension (specifically the Kerr-NUT-(A)dS class of Ref. \citep{Chen:2006xh}).

%%%%%%%%%%%%%%%%%%%%%%%%%%%%%%%%%%%%%%%%%%%%%%%%%%%%%%%%%%%%%
\mytitle{Acknowledgements}
The work on this paper was supported by the Czech Science Foundation grant 26-23696S.

\mytitle{Data availability}
The computations in the example spacetimes are reproduced in the  \textit{Mathematica} notebook \texttt{SpinSepExamples.nb} in the Supplemental Material.
%%%%%%%%%%%%%%%%%%%%%%%%%%%%%%%%%%%%%%%%%%%%%%%%%%%%%%%%%%%%%
\bibliography{refs}
%%%%%%%%%%%%%%%%%%%%%%%%%%%%%%%%%%%%%%%%%%%%%%%%%%%%%%%%%%%%%

%%%%%%%%%%%%%%%%%%%%%%%%%%%%%%%%%%%%%%%%%%%%%%%%%%%%%%%%%%%%%
% End matter/Appendix (replaces *online only* Supplemental material)
%%%%%%%%%%%%%%%%%%%%%%%%%%%%%%%%%%%%%%%%%%%%%%%%%%%%%%%%%%%%%
\appendix
\section*{End matter}

%%%%%%%%%%%%%%%%%%%%%%%%%%%%%%%%%%%%%%%%%%%%%%%%%%%%%%%%%%%%%
\subsection{Canonical transformation and phase-space Hamiltonian}
\label{sec:supp_canon}
%%%%%%%%%%%%%%%%%%%%%%%%%%%%%%%%%%%%%%%%%%%%%%%%%%%%%%%%%%%%%
The type-II generating function of the main text, $G=\tilde x^\mu p_\mu+\tilde\Lambda^{C}{}_{\hat A}\,L^{B}{}_C(\tilde x,p)\,\Pi_{B}{}^{\hat A}$ with $L^A{}_B$ a phase-space-dependent Lorentz matrix, gives $\Lambda^{B}{}_{\hat A}=L^{B}{}_C\,\tilde\Lambda^{C}{}_{\hat A}$ and $\tilde\Pi_{B}{}^{\hat A}=L^{C}{}_B\,\Pi_{C}{}^{\hat A}$, hence the tetrad rotation $S_{AB}=\tilde S_{CD}\,L_{A}{}^{C}L_{B}{}^{D}$ together with the momentum and coordinate shifts quoted there. Equivalently, in terms of the covariant momentum $P_\mu$ ($g^{\mu\nu}P_\mu P_\nu=-\mathcal{M}^2$) and the adapted tetrad $e^\mu_A(x,p)$,
\begin{align}
	\tilde p_\mu = P_\mu + \tfrac{1}{2}\,e_{\nu A;\mu}\,e^\nu_{B}\,S^{AB}\,, \;
	x^\mu = \tilde x^\mu - \tfrac{1}{2}\,\pdv{e_{\nu A}}{p_\mu}\,e^\nu_{B}\,S^{AB}. \label{eq:supp_covshift}
\end{align}
Writing $P_\mu=p_\mu-\tfrac12\,\omega_{\mu AB}S^{AB}$ and $x^\mu=\tilde x^\mu+\tfrac12\,\alpha^\mu{}_{AB}S^{AB}$ with $\omega_{\mu AB}=e^\nu_{A;\mu}e_{B\nu}$ and $\alpha^\mu{}_{AB}=\pdv{e_{\nu A}}{p_\mu}e^\nu_{B}$, inserting into $H=\tfrac{1}{2\mathcal{M}}g^{\mu\nu}(\tilde x)P_\mu P_\nu$, and using $\tfrac12 g^{\mu\nu}{}_{,\rho}=-\Gamma^{(\mu\nu)}{}_{\rho}$, collects the two $\mathcal O(S)$ terms into the phase-space covariant derivative $\mathcal{D}_\mu$ of the main text and reproduces Eq.~\eqref{eq:Hmin}. These relations are exact; since the shifts are $\mathcal O(S)$, the transform and its inverse follow by a single iteration from the identity. The derivative $\mathcal{D}_\mu$ is the natural connection along the phase-space flow: $\tfrac1{\mathcal M}\,p_\nu g^{\mu\nu}\mathcal{D}_\mu v^\alpha=0$ reproduces ordinary parallel transport $Dv^\alpha/\d\tau=0$ once $\d x^\mu/\d\tau=g^{\mu\nu}p_\nu/\mathcal{M}$ is used.

%%%%%%%%%%%%%%%%%%%%%%%%%%%%%%%%%%%%%%%%%%%%%%%%%%%%%%%%%%%%%
\subsection{Momentum frame, Tulczyjew--Dixon constraint, and Newton--Wigner variables}
\label{sec:supp_nw}
%%%%%%%%%%%%%%%%%%%%%%%%%%%%%%%%%%%%%%%%%%%%%%%%%%%%%%%%%%%%%
Choosing the time leg along the momentum, $e^\mu_0=u^\mu\equiv p^\mu/\sqrt{-p_\nu p^\nu}$, and building the spatial legs by the (Wigner) boost of a background tetrad $\tilde e^\mu_A(x)$ into the momentum rest frame,
\begin{align}
	e^\mu_{I} = \tilde e^\mu_{I} + \frac{u_\nu\,\tilde e^\nu_{I}}{1 - u^\nu \tilde e_{\nu 0}}\bigl(u^\mu + \tilde e^{\mu}_{0}\bigr)\,, \; I=1,\dots,N-1\,, \label{eq:supp_boost}
\end{align}
renders the Tulczyjew--Dixon supplementary condition algebraically trivial, $S^{\mu\nu}P_\nu=0\Leftrightarrow S_{0B}=0+\mathcal O(S^2)$. (The leg \eqref{eq:supp_boost} is orthonormal off-shell and reduces to $\tilde e^\mu_I$ as $u^\mu\to\tilde e^\mu_0$.) The boost components $S_{0I}$ are second class, $\{S_{0I},S_{0J}\}=-S_{IJ}$, whereas the rotation sector closes on the $\mathfrak{so}(N-1)$ algebra of the main text with no $S_{0B}$ appearing; hence on the surface $S_{0B}=0$ the Dirac bracket of the $S_{IJ}$ coincides with their Poisson bracket, covered in $N=4$ by the single canonical pair $(\phi,\pi_\phi)$ given there.

With the choice \eqref{eq:supp_boost}, the connection term in \eqref{eq:Hred} becomes
\begin{align}
	\mathcal{D}_\nu e^\kappa_I e_{J\kappa} = \tilde e^\kappa_{I;\nu}\tilde e_{J\kappa}
	- \frac{2}{1-u^\lambda\tilde e_{\lambda 0}}\,\tilde e_{\kappa 0;\nu}\,\tilde e^\kappa_{[I} u_{J]}\,,
\end{align}
and reducing $-p_t$ on the mass shell gives the coordinate-time Hamiltonian $H_t=H_0+H_S$ with
\begin{align}
	& H_0 = \sqrt{\frac{g^{ij}}{-g^{tt}} p_i p_j + \frac{\mathcal{M}^2}{-g^{tt}} + \Bigl(\frac{g^{ti}}{g^{tt}}p_i\Bigr)^2} - \frac{g^{ti}}{-g^{tt}} p_i \,, \\
	& H_S = -\frac{U^\nu}{U^0}\Bigl(\tilde e^\kappa_{I;\nu}\tilde e_{J\kappa} - \frac{2}{1-u^\lambda\tilde e_{\lambda 0}}\,\tilde e_{\kappa 0;\nu}\,\tilde e^\kappa_{[I} U_{J]}\Bigr)\frac{S^{IJ}}{2}\,, 
    \\
    & U_\mu\equiv(-H_0,\,p_i)/\mathcal{M}\,.
\end{align}
where the two entries in $U_\mu$ denote the time and spatial components.
% \vs{Přidal bych vysvětlení notace pro $U_\mu$ (jako je za rovnicí (62))} 
Comparing with the Dirac/Newton--Wigner reduction of \citet{Barausse:2009aa}, the two Hamiltonians are \emph{identical} up to the relabelling $S_I=\tfrac12\epsilon_{IJK}S^{JK}$ between the spin tensor and the spin vector measured in the momentum rest frame.

Concretely, the dictionary is the following. Their reference tetrad, particle mass $m=\sqrt{-p_\mu p^\mu}$, and canonical four-momentum $(-H,P_i)$ [their Eq.~(3.27)] are our $\tilde e_A$ (their timelike label $T$ being our $0$), $\mathcal{M}$, and $p_\mu$, while their kinetic momentum $p_\mu$ is our covariant $P_\mu$---i.e.\ the symbols $p$ and $P$ are interchanged between the two papers. Our momentum-frame condition $S_{0B}=0$ is precisely their generalized Newton--Wigner SSC $S^{\mu\nu}\omega_\nu=0$ with $\omega_\mu=p_\mu-m\,\tilde e^{T}_\mu$ [their Eqs.~(4.6)--(4.7)], the minimal boost~\eqref{eq:supp_boost} being the transformation that maps one supplementary condition into the other. Consequently $H_0$ equals their non-spinning Hamiltonian $\bar H_{\mathrm{NS}}=\beta^iP_i+\alpha\sqrt{m^2+\gamma^{ij}P_iP_j}$, with $\alpha=1/\sqrt{-g^{tt}}$, $\beta^i=g^{ti}/g^{tt}$, $\gamma^{ij}=g^{ij}-g^{ti}g^{tj}/g^{tt}$ [their Eq.~(4.45)], and $H_S$ reproduces the spin coupling of their Eq.~(4.44) once the spin tensor is traded for the vector $S_I=\tfrac12\epsilon_{IJK}S^{JK}$ [their Eq.~(4.26)]; the agreement may be checked term by term, e.g.\ for the spherically symmetric Hamiltonian of their Eq.~(5.18). 

%%%%%%%%%%%%%%%%%%%%%%%%%%%%%%%%%%%%%%%%%%%%%%%%%%%%%%%%%%%%%
\subsection{Metric functions in Kerr-NUT-(A)dS spacetimes}
%%%%%%%%%%%%%%%%%%%%%%%%%%%%%%%%%%%%%%%%%%%%%%%%%%%%%%%%%%%%%
The metric functions appearing in equation \eqref{eq:CarterMetric} are \citep{griffiths2009exact}
\begin{align}
		\Delta_r      &= \bigl(a^2 - n^2 + e^2 + g^2\bigr) - 2Mr + r^2
		\notag\\
		&\quad
		- \Lambda\!\left[(a^2-n^2)\,n^2 + \bigl(\tfrac{a^2}{3}+2n^2\bigr)r^2 + \tfrac{r^4}{3}\right], \\[4pt]
		\Delta_\theta &= 1 + \tfrac{\Lambda}{3}\,a\cos\theta\,(4n + a\cos\theta).
	\end{align}
The parameters are black hole mass $M$, spin $a$, NUT charge $n$,
cosmological constant $\Lambda$, and electric/magnetic charges $e,g$. With the substitution of these functions, the metric fulfills source-free Einstein-Maxwell equations.
However, we repeat that the results in the main text apply even if the functions are ``off-shell'', that is, the form of $\Delta_r(r)$ and $\Delta_\theta(\theta)$ is such that the Einstein equations are not necessarily fulfilled. 

%%%%%%%%%%%%%%%%%%%%%%%%%%%%%%%%%%%%%%%%%%%%%%%%%%%%%%%%%%%%%
\subsection{Separation functions for Kerr--NUT--(A)dS geodesics}
%%%%%%%%%%%%%%%%%%%%%%%%%%%%%%%%%%%%%%%%%%%%%%%%%%%%%%%%%%%%%
The geodesic Hamilton--Jacobi equation \eqref{eq:HJgeo} in the Kerr--NUT--(A)dS metric separates with the characteristic function
\begin{align}
\begin{split}
	&W_{(\mathrm g)} = -E\,(t-t_0) + L\,(\varphi-\varphi_0)
	 \\& \pm \int \frac{\sqrt{R(r;E,L,K)}}{\Delta_r}\,\d r
	 \pm \int \frac{\sqrt{\Theta(\theta;E,L,K)}}{\Delta_\theta}\,\d\theta\,,
\end{split}
\end{align}
where $E=-W_{(\mathrm g),t}$ and $L=W_{(\mathrm g),\varphi}$ are the conserved energy and azimuthal angular momentum, and $K=K_{\mu\nu}u^\mu u^\nu$ (with $u_\mu=W_{(\mathrm g),\mu}$, $u_\mu u^\mu=-1$) is the Carter constant associated with the Killing tensor $K_{\mu\nu}=Y_\mu{}^\kappa Y_{\nu\kappa}$. The radial and polar functions read
\begin{align}
    \begin{split}
	& R(r;E,L,K) = \bigl[E\bigl(r^2+(a+n)^2\bigr)-aL\bigr]^2 
    \\&
    \phantom{R(r;E,L,K) =} - \Delta_r\bigl(r^2 + K\bigr)\,, 
    \end{split}
    \\
    \begin{split}
        & \Theta(\theta;E,L,K) = \Delta_\theta\bigl[K-(n+a\cos\theta)^2\bigr]
	   \\ & \qquad - \frac{\bigl[L-\bigl(a\sin^2\!\theta+2n(1-\cos\theta)\bigr)E\bigr]^2}{\sin^2\!\theta}\,.
    \end{split}
\end{align}
This implies $\d r/\d\tau=\pm\sqrt{R}/\rho^2$, $\d\theta/\d\tau=\pm\sqrt{\Theta}/\rho^2$, i.e.\ $\d r/\d\lambda=\pm\sqrt{R}$, $\d\theta/\d\lambda=\pm\sqrt{\Theta}$ for a Mino-like parametrization $\d\tau/\d\lambda=\rho^2$. In the Kerr limit ($n=\Lambda=0$, $\Delta_\theta=1$), the functions $R,\Theta$ reduce to the standard Carter forms. Throughout the main text we adopt the proper-time normalization $u_\mu u^\mu=-1$ ($\mathcal{M}=1$), so that $\omega_{12}$, $\psi$, and $W_{(1)}$ carry no explicit factors of $\mathcal{M}$; the canonical (full-mass) forms follow from the substitution $u_\mu\to W_{(0),\mu}/\mathcal{M}$, with $E,L,K$ then the full constants $-p_t,\,p_\varphi,\,K^{\mu\nu}p_\mu p_\nu$ ($p_\mu=W_{(0),\mu}$, $\mathcal{M}^2=-p_\mu p^\mu$), the precession denominators becoming $K+\mathcal{M}^2 r^2$ and $K-\mathcal{M}^2(n+a\cos\theta)^2$ and the integrand of $\psi$ acquiring an overall factor $\mathcal{M}$.

%%%%%%%%%%%%%%%%%%%%%%%%%%%%%%%%%%%%%%%%%%%%%%%%%%%%%%%%%%%%%
\subsection{Worldline-adapted tetrad in Kerr--NUT--(A)dS}
%%%%%%%%%%%%%%%%%%%%%%%%%%%%%%%%%%%%%%%%%%%%%%%%%%%%%%%%%%%%%
\label{sec:supp_kerrtet}
With the Killing--Yano tensor $Y_{\mu\nu}$ and the Killing tensor $K_{\mu\nu}=Y_\mu{}^\kappa Y_{\nu\kappa}$ of the main text, the Marck-type tetrad referred to there is
\begin{align}
	& e_{0 \mu} = \frac{u_{\mu}}{\sqrt{-u_\kappa u^\kappa}} \,,
	\\
	& e_{3\mu} = \frac{1}{\sqrt{K}}\,Y^{\nu}{}_{\mu}u_\nu \,,
	\\
	& e_{1\mu} = \frac{1}{N_{(1)}} \left(K_{\mu \nu} + \frac{K  g_{\mu \nu}}{-u_\kappa u^\kappa} \right) u^\nu \,,
	\\
	& e_{2\mu} = \epsilon_{\mu\nu\kappa\lambda}\, e_{0}^{\nu} e_{1}^{\kappa} e_{3}^{\lambda} \,,
	\\
	& K \equiv K^{\mu \nu} u_\mu u_\nu \,,\quad N_{(1)}^2\equiv K_{\mu \nu}K^{\nu}{}_{\kappa} u^\mu u^\kappa - \frac{K^2}{u_\kappa u^\kappa}\,.
\end{align}
Unlike the construction of Ref.~\cite{Witzany:2019}, this tetrad is orthonormal off-shell and regular through the point $n+a\cos\theta=0$ where $Y_{\mu\nu}$ degenerates. By the geodesic equation and the defining relations of $Y_{\mu\nu}$, the legs $e_{0\mu}$ and $e_{3\mu}$ are parallel transported along geodesics, leaving the single precession component $\omega_{12}$ quoted in the main text.

%%%%%%%%%%%%%%%%%%%%%%%%%%%%%%%%%%%%%%%%%%%%%%%%%%%%%%%%%%%%%
\subsection{Transverse legs of the plane-wave tetrad}
%%%%%%%%%%%%%%%%%%%%%%%%%%%%%%%%%%%%%%%%%%%%%%%%%%%%%%%%%%%%%
\label{sec:supp_pw}
For the plane-wave example of the main text, alongside the timelike leg $e_0^\mu=g^{\mu\nu}u_\nu/\sqrt{-u_\kappa u^\kappa}$ and the longitudinal leg $e_3^\mu$ built from the covariantly constant null vector, the two transverse legs along the principal directions of the wave are
\begin{align}
	e_1^\mu \partial_\mu &= e^{-\beta-\gamma}\Big[\cos\alpha\,\big(\partial_x - \tfrac{u_x}{u_V}\partial_V\big) + \sin\alpha\,\big(\partial_y - \tfrac{u_y}{u_V}\partial_V\big)\Big],\\
	e_2^\mu \partial_\mu &= e^{-\beta+\gamma}\Big[{-}\sin\alpha\,\big(\partial_x - \tfrac{u_x}{u_V}\partial_V\big) + \cos\alpha\,\big(\partial_y - \tfrac{u_y}{u_V}\partial_V\big)\Big].
\end{align}
They are orthonormal off-shell and rotate in the transverse plane at the rate $\d\alpha/\d U$, which produces the single nonzero precession component $\omega_{12}=p_V\cosh(2\gamma)\,\d\alpha/\d U$ quoted in the main text.

%%%%%%%%%%%%%%%%%%%%%%%%%%%%%%%%%%%%%%%%%%%%%%%%%%%%%%%%%%%%%
\subsection{Worldline-adapted tetrad for FLRW spacetimes}
\label{sec:supp_flrw}
%%%%%%%%%%%%%%%%%%%%%%%%%%%%%%%%%%%%%%%%%%%%%%%%%%%%%%%%%%%%%
For the FLRW example of the main text the adapted tetrad is constructed as follows. The timelike leg is taken along the four-velocity and the first spatial leg along the comoving spatial velocity,
\begin{align}
	& e^\mu_0 = \frac{g^{\mu\nu}u_\nu}{\sqrt{-u_\kappa u^\kappa}}\,,
	\\
	& e^\mu_1 = \frac{1}{a\sqrt{-u_\kappa u^\kappa}}\left(\sqrt{\gamma^{kl}u_k u_l}\,,\ \frac{-u_t\,\gamma^{ij}u_j}{\sqrt{\gamma^{kl}u_k u_l}}\right)\,,
\end{align}
where the two entries denote the time and spatial ($x^i$) components. The remaining freedom in the transverse plane is fixed using the gradient of the comoving geodesic distance from $\chi=0$, $\rho_\mu=\delta_\mu^\chi$,
\begin{align}
	& e^\mu_2 = \mathcal{N}\,\epsilon^{\mu\nu\kappa\lambda}\,e_{0\nu}\,e_{1\kappa}\,\rho_\lambda\,,
	\\
	& e^\mu_3 = \mathcal{N}\Big(\rho^\mu + (\rho_\nu e_0^\nu)\,e_0^\mu - (\rho_\nu e_1^\nu)\,e_1^\mu\Big)\,,
	\\
	& \mathcal{N} \equiv \frac{a}{\sqrt{1 - u_\rho^2/(\gamma^{kl}u_k u_l)}}\,,\quad u_\rho\equiv\gamma^{ij}\rho_i u_j \,,
\end{align}
with $\epsilon^{\mu\nu\kappa\lambda}$ the Levi-Civita pseudotensor. The construction degenerates for purely radial geodesics $u_i\propto\rho_i$, where one simply picks a reference point other than $\chi=0$. With this choice \emph{every} spatial leg is parallel transported along the geodesics, so the transport matrix vanishes identically, $\omega_{IJ}=0$; the precession angles are trivial and the main-text spin correction reduces to $W_{(1)}=\sigma\phi$ with $\sigma=S_{12}$.

\end{document}